\documentclass{cjaa}                   
\usepackage{graphicx}                   
\input{epsf.sty}                        
\begin{document}

\title{Toward an understanding of the periastron puzzle of PSR B1259$-$63}

   \volnopage{Vol.0 (200x) No.0, 000--000}      
   \setcounter{page}{1}          

   \author{Ren-Xin Xu
      }
   \offprints{Ren-Xin Xu}                   

   \institute{Astronomy Department, School of Physics, Peking University, Beijing 100871, China\\
             \email{r.x.xu@pku.edu.cn}}

\date{Received~~~~~~~~~~~~~~~~~; ~~accepted}

   \abstract{
Efforts are made to understand the timing behaviors (e.g., the
jumps of the projected pulsar semimajor axis in the periastron
passages) observed via 13-year timing for PSR B1259-63.
In the first, planet-like objects are suggested to orbit around
the Be star, which may gravitationally perturb the (probably low
mass) pulsar when it passes through periastron.
Second, an accretion disk should exist outside the pulsar's light
cylinder, which creates a spindown torque on the pulsar due to the
propeller effect. The observed negative braking index and the
discrepant timing residuals close to periastron could be
relevant to the existence of the disk with varying accretion
rates.
Third, a speculation is presented that the accretion rate may increase on
a long timescale in order to explain the negative braking index.
   }

   \authorrunning{Xu}            
   \titlerunning{Toward an understanding of the periastron puzzle}  

   \maketitle


\section{Introduction}

Pulsars are clocks with great precision, with which one can probe
astrophysical processes. The pulsar clock technique has been very
successful in studies of dynamical systems (both general
relativistic and Newtonian) and pulsar interiors (glitches).
Thirteen years of timing data have been obtained  for the unique
pulsar PSR B1259-63, which has a massive companion with a circumstellar
disk  (Wang et al. 2004). Unfortunately,  as concluded by
Wang et al. (2004), a plausible mechanism to account for the
observed timing behavior still has not been offered.

PSR B1259-63 was discovered in a survey of the Galactic plane by
Johnston et al. (1992), but determining its timing properties has
not been easy.
Manchester et al. (1995) fitted the timing data by introducing
$\Delta P/P\sim 10^{-9}$ at each periastron, which they attributed to
propeller torque spindown.
Wex et al. (1998) presented a timing model with the inclusion of
changing $\dot \omega$ and $\dot x = {\rm d/d}t [a_p \sin i]$
terms, which still failed to accurately describe the 13-year
dataset (Wang et al 2004).
{\it{Jumps}} in projected semimajor axis $\Delta x = \Delta(a_p
\sin i)$ at periastron were introduced by Wang et al. (2004). The
rms residual in the orbit-jump model is only about half that in
the $\nu$-$\dot\nu$-jump model, even though the free parameters in
the later model are four more than in the former one.

PSR B1259-63 may accrete from the dense circumstellar environment
when it passes though periastron.  Observations of the pulsar
enable  one to probe the circumstellar medium of an accreting
pulsar.
Recent studies show that propeller disks around such compact stars
may play an important role for various astrophysical objects,
including anomalous X-ray pulsars and soft $\gamma$-ray repeaters
(AXP/SGRs, e.g., Chatterjee et al. 2000) and compact center
objects (CCOs, Xu et al. 2003).
Recently, evidence for a fossil disk around an anomalous X-ray
pulsars, 4U 0142+61, has been proposed (Wang et al. 2006) through
detecting emission at two mid-infrared bands ($4.5 \mu$m and $8.0
\mu$m)\footnote{%
However, no emission is detected at $24\mu$m and $70\mu$m, with
limits of 0.05 mJy and of 1.5 mJy, respectively (Bryden et al.
2006). These two observations are consistent with Xu (2005), where
at least part of the millisecond pulsars are suggested to be of
supernova origin via accretion-induced collapses.
}. %
PSR B1259-63, as an ideal ``laboratory'', can be studied in great
detail, and is the {\em only} known case of  a pulsar's disk with
material replenishment (from the wind-disc of the Be star).

\section{The $x$-jump scenarios proposed}

The inclination of the system to our line of sight angle
$i=\sin^{-1}\{[f(M_p)(M_p+M_c)^2]^{1/3}/M_c\}$ is $\sim 35^{\rm
o}$ ($i=32^{\rm o}\sim 36^{\rm o}$ for $M_p/M_\odot=0\sim 1.4$)
for the companion Be star with mass $M_c=10M_\odot$ and the mass
function $f(M_p)=1.53M_\odot$.
The total orbital angular momentum $L$ for a Keplerian orbit reads
\begin{equation}
\begin{array}{lll}
L & \simeq & M_pM_c\sqrt{G(1-e^2)a_p\over M_p+M_c}=M_p\sqrt{G(1-e^2)M_cx_p\over \sin i}\\
& \sim & 3\times 10^{53}M_{p1}M_{c10}^{1/2}~{\rm g~cm^2/s},%
\end{array}
\label{L}
\end{equation}
for the PSR B1259$-$63 system when $M_p\ll M_c$ ($i\sim 35^{\rm
o}$), where $a_p$ is the semimajor axis and the projected pulsar
semimajor axis $x_p=a_p\sin i$, $M_p=M_\odot M_{p1}$,
$M_c=10M_\odot M_{c10}$ (``$p$'' and  ``$c$'' denote the $p$ulsar
and the $c$ompany Be star, respectively).

The spin angular momentum $S_c$ of the
Be star can be approximated as
\begin{equation}
\begin{array}{lll}
S_c & \sim & 0.4M_cR_c^2\Omega_c\sim 0.4\sqrt{G}fM_c^{3/2}R_c^{1/2}\\
& \sim & 8\times 10^{52}f_{0.5}M_{c10}^{3/2}R_{c6}^{1/2}~{\rm g~cm^2/s},%
\end{array}
\label{S}
\end{equation}
where the Be star is rotating at $f$ times of the centrifugal
breakup frequency $\Omega_k\simeq \sqrt{GM_c/R_c^3}$, $f=0.5
f_{0.5}$, the Be star radius $R_c=6R_\odot R_{c6}$. It is evident
that the magnitude of $L$ and that of $S_c$ are comparable. It
would not be difficult to change ${\bf L}$ (while keeping $|{\bf
L}|$ almost the same) via occasional spin-orbit coupling (i.e.,
transferring orbital to spin angular momentum,
 or vice versa).

As noted by Wang et al. (2004),  it  is not likely that the
$x$-jumps are changes in semimajor axis $a_p$.  Rather, assuming a
constant $a_p$, then a $\Delta x = \Delta (a_p \sin i)$ requires
an inclination angle change per periastron passage
\begin{equation}
\Delta i = (\Delta x_p/x_p) \tan i\sim 3^{\rm o}\times
10^{-4}\Delta x_{p10},
\label{Di}
\end{equation}
where $\Delta x_p=10{\rm ms}~c~\Delta x_{p10}$, with $c$ the speed
of light.
If $|{\bf L}|$ is the same, the orbital angular momentum change $\Delta L$,
which is nearly perpendicular to {\bf L}, is
\begin{equation}
\Delta L \simeq 5.4\times 10^{-6}\Delta x_{p10}L\sim 10^{48}\Delta
x_{10}M_{p1}M_{c10}^{1/2}~{\rm g~cm^2/s},
\label{DL}
\end{equation}
which shows $\Delta L \ll L$.

No $\Delta i$ can occur if the mass distribution is symmetric with
respect to the orbital plane. Note that $\Delta i$ should have
only one sign as long as the asymmetric pattern of
mass-distribution is almost the same (e.g., the case in which an
asymmetric pattern originates only from an angle between the spin
axis of Be star and the orbit normal).
One must therefore find mechanisms that can cause an asymmetric, time-variable
mass distribution.

{\em Planets around the Be star?}
Neither observational nor theoretical conclusions of massive star
formation and evolution are certain. Nevertheless, a general
view for B-spectral class stars could be as follows: After
gravitational collapse of a molecular cloud, a massive star may
form inside the cloud, with a dense, dusty circumstellar disk and
strong bipolar outflow. The star could then become a so-called
young Herbig Be star if it is still embedded in a nebula (Fuente
et al. 2003).  It would then evolve to be a normal Be star (like SS
2883, the companion to PSR B1259-63, without a nebula).  The
outflow is weak (or there is a tenuous polar wind), and its disk
becomes less and
 less massive. Finally, the star becomes a B star when the disk
disappears (like the companion of PSR J0045-7319).

One may speculate that planet-like objects (planets,
proto-planets, or planetesimals) could form in the dusty disks of
a Be star after the first supernova of binary, with highly
eccentric orbits. The number of such objects and thus the total
mass decrease with time due to their being captured,
disintegrated, or evaporated by their hot Be stars (or via the
interaction with the disk-wind).
Evidence for planetesimal (with asteroidal size) infall
onto a very young ($\sim 10^5$ years) Herbig Be star
LkH$_\alpha$234 has recently been reported (Chakraborty, Ge \& Mahadevan 2004).
More than 100 extra-solar planets, with masses $\sim (1-10)M_{\rm
Jupiter}$, have been discovered by radial-velocity
surveys (Tremaine \& Zakamska 2004).
Alternatively, planets formed before the first supernova could
probably be residual if the supernova occurs far way its company
at a very highly eccentric orbits.

There might be many planet-like objects (planets, proto-planets,
planetesimals, or even brown dwarfs) orbit around the Be star; but
the exact number could hardly obtained by observation or
calculation.
PSR B1259-63 would be perturbed by the gravity of one or more such
planet-like bodies near its periastron if they orbit the Be star
with their semimajor axis being much smaller than the pulsar's,
$a_p$.
However, if the pulsar is though to have a conventional, normal
mass $\sim 1.4M_\odot$, the perturbation should be negligible.
Nonetheless, pulsars could be quark stars having masses
much smaller than $\sim M_\odot$ (Xu 2005).  Hence the perturbation
might be strong enough for the observed  $x$-jumps if PSR B1259-63
is a low-mass quark star.
The orbital angular momentum of a planet around the Be star is
$L_{\rm planet}\sim 2\times 10^{47}M_{J}a_{100}^{1/2}~{\rm
g~cm^2/s}$, where the planet mass is $M_J\times M_{\rm Jupiter}$
and its semimajor axis is $a_{100}\times 100R_\odot$.
From Eq.(\ref{DL}), one sees $\Delta L<L_{\rm planet}$ if $M_J\ll
M_p<0.2M_Ja_{100}^{1/2}\Delta x_{10}^{-1}M_\odot$, and a
significant $x$-jump of a low-mass quark star is possible through
the gravitational perturbation of planet-like objects.
An encounter of a planet and the pulsar per periastron passage
would produce (1) a large exchange of angular momentum between the
orbits (thus an $x$-jump) and (2) a large change in the orbital
elements of the planet (thus a higher eccentricity of the planet).
The migration inwards due to the interaction of the planets with
the disk would increase the possibility of the encounters and may
decrease the eccentricity, but the pulsar might in turn ``eject''
planets after encounters.

It may be hard to believe that planets would exist around a Be
star since it is conventionally thought that planets form by
accretion over $\sim 10^7$ years. Nevertheless, planet formation
could be sped up due to fragmentation of an unstable disk (Boss
2003).
Another possibility that can not be ruled out, is that the planets
could be strange planets (i.e., strange quark matter with planet
masses) which were born during the first supernova explosion.
This kind of planet can not evaporate, and thus would exist for a long
time.

{\em Asymmetric mass distribution of the Be star?}
Besides the suggestion above, two more scenarios are speculated
for $x$-jumps.
The periastron separation between the pulsar and the Be  star is
$r_p \sim 24 R_c\gg R_c$.
If the Be star has stellar oscillations leading to multipoles of
mass distribution, then the torque acting on the pulsar may result
in an $x$-jump per periastron passage. As no torque is related to
the monopole term, the quadrupole one would be important for
$x$-jumps.
It worth noting that the oscillations could be enhanced if they
are excited resonantly (e.g., when the ratio of the orbital period
to an oscillation period is an integer) by the orbiting pulsar
(Witte \& Savonije 1999).
Chaotic orbital dynamics near the periastron could also be
possible in this case (Mardling 1995), which would result in a
random transfer of angular momenta between comparable values of
$L$ and $S_c$ (Eq.(\ref{L}-\ref{S})).
Additional study of the structure and evolution of massive stars,
especially their oscillation behaviors, might further elucidate
the periastron puzzle in this scenario.

{\em Asymmetric mass distribution of the disk of SS 2883?}
The existence of the Be star disk is an obvious asymmetry with respect
to the orbital plane. But this can not cause sign-changed
$x$-jumps if the disk is almost homogeneous.
However, if a density-wave pattern can be excited (e.g., by the
pulsar or planets) in the disk, like the case of Saturn's ring
structure affected by the satellite Mimas, then angular momentum may
transfer between the orbiting pulsar and the disk in the
periastron passages, which might result in $x$-jumps. But it seems
difficult to limit the duration of this effect (if it exists) only to the period of
 the pulsar passing near the Be star's disk.

\section{Propellers in a radio pulsar phase?}

One strange feature of this system is that the braking index is
about $-37$ after 13 years of  timing of PSR B1259-63, while the
indices are between 1.4 and 2.9 ($\la 3$) for 6 pulsars, with
great certainty (Xu \& Qiao 2001; Livingstone et al. 2006).
It is true that the data-fitting is improved significantly when
$x$-jumps at periastron passages and a pulsar glitch near MJD
50691 are included, whereas a few residual points close to the
periastron are still discrepant systematically.
This could be due to errors in the DM correction (Wang et al.
2004). An alternative explanation presented below for the unusual
braking index as well as the timing residuals is that the pulsar
may capture SS 2883's circumstellar matter to form a pulsar disk
which results in a torque being antiparallel to the pulsar's spin
(the viscosity timescale of accretion disk could be order of
years).

By observing the un-pulsed radio emission at four different bands,
Johnston et al. (2005) conclude that the pulsar passes through the
dense circumstellar disk of the Be star just before and just after
periastron.
It iseems intuitive that accretion would occur when the pulsar passes
through the disk.
Accretion of in-falling matter with negligible angular momentum
was considered by Kochanek (1993) and Manchester et al. (1995),
in which the pulsar loses its angular momentum in order for the
in-falling matter to obtain enough angular momentum to escape.
Besides spinning down the pulsar, the no-angular-momentum accretion may
also affect pulsar orbit parameters through frictional drag (Wex
et al. 1998).
However, if the material captured by the pulsar carries significant angular
momentum, disk accretion around the pulsar occurs.
To make this more precise, one should calculate the circularization
radius to see if the accreted matter has sufficient angular
momentum for disk formation.

{\em Accretion disk formation around PSR B1259-63.}
The cylindrical radius, $r_{\rm acc}$, within which a pulsar with
mass $M_p$ can gravitationally capture material with a relative
velocity $v_{\rm rel}$ can be estimated as (e.g., Lipunov 1992)
\begin{equation}
r_{\rm acc}\simeq GM_p/v_{\rm rel}^2\sim 1.3\times
10^{12}M_{p1}v_{\rm rel100}^{-2}~{\rm cm},
\label{racc}
\end{equation}
where $v_{\rm rel100}=v_{\rm rel}\times 100$ km/s. Note that the
radius of PSR B1259-63's light cylinder is only $r_{\rm
lc}=cP/(2\pi)=2.3\times 10^8$ cm.
If the accreted material is from the Be star's wind with velocity $v_w\sim
500$ km/s, the circularization radius (Frank et al. 2002) in this
case is only $R_{\rm circ}^{\rm w}\simeq G^3M_p^3\Omega_{\rm
orb}^2/v_w^8\sim 200M_{p1}^3$ cm, which is much smaller than the pulsar's
radius, where $\Omega_{\rm orb}=2\pi/P_{\rm orb}$.
Therefore a disk around the pulsar is not likely to form by accretion from the Be
star's wind\footnote{
Even in this case of negligible momentum of accreting wind, a disk
might form due to the interaction between the infalling medium and
the rapidly spinning pulsar's magnetosphere, by which the
rotational angular momentum of the pulsar is transferred to the
accreted material.
}. %

However, if the gas in the Be star's disk moves in a Keplerian orbit
and the pulsar has an orbital velocity near periastron ($r_p \sim
24R_c\sim 10^{13}$ cm) of $v_{\rm peria}=\sqrt{GM_c(2/r_p-\sin
35^{\rm o}/x_p)}\sim 157$ km/s, the accretion radius $r_{\rm acc}$
is different for material inside and outside periastron.
The Keplerian velocity  periastron is
$v_k(r_p)=\sqrt{GM_c/r_p}\sim 115$ km/s, but the sonic speed of
the line-emission disk with temperature $\sim 10^4$ K is only
$v_s\sim 10$ km/s $\ll v_k$. We will then neglect the pressure of
disk material in the following consideration.
Assuming that the pulsar orbits in the same direction as  the
circumstellar medium around the Be star, for $M_p=1M_\odot$, the
Keplerian velocity at the accretion boundary outside or inside the
periastron is $v_k=\sqrt{GM_c/(r_p \pm r_{\rm acc})}$.  Using
Eq.(\ref{racc}) and $v_{\rm rel}\simeq v_k-v_{\rm
peria}$, one can find $r_{\rm acc}\sim 3.8\times 10^{12}$ cm and
$\sim 6.7 \times 10^{12}$ cm for outside and inside the
periastron, respectively.
The specific angular momentum\footnote{%
Note that $v_k({\rm outside})\simeq 98$ km/s $< v_{\rm peria} <
v_k({\rm inside})\simeq 201$ km/s in this case.
} %
is then order of $l\sim r_{\rm acc}v_{\rm rel}$, and the
circularization radius is then $R_{\rm circ}^{\rm
d}=l^2/(GM_p)\sim 10^{11}$ cm for $r_{\rm acc}\sim 10^{12}$ cm,
being much larger than $r_{\rm m}$ (Fig.1, $r_{\rm lc}\sim
10^8$cm).
One can still have $R_{\rm circ}^{\rm d}\sim 10^{11}\gg r_{\rm
lc}$ even for a small pulsar mass of  $M_p=10^{-3}M_\odot$
($r_{\rm acc}\sim 7.7\times 10^9$ cm and $\sim 7.2 \times 10^{12}$
cm for outside and inside the periastron, respectively, in this
case).
Therefore an accretion disk can form around the pulsar if the
circumstellar disk of the Be star is Keplerian, even without the
inclusion of magnetospheric interaction\footnote{%
This interaction transfers angular momentum from pulsar to
accretion matter, and thus favors disk formation.
}, %
since the circularization radii are usually much larger than the
magnetospheric ones.

{\em Propeller torque and the pulsar's spindown.}
For simplicity, we assume that the angular momentum of the
pulsar's disk is parallel to the spin axis of PSR B1259-63.
If the circumstellar matter around the Be star is a hydrogen
plasma, the mass density would be (Wex et al. 1998)
%
$\rho(r)\simeq 7.5\times 10^{-12}(r/R_c)^{-4.2}~{\rm g/cm^3}$,
%
%
where the Be star's radius $R_c\sim 6R_\odot$.
When the pulsar passes through the dense disk near periastron, the
accretion rate could be
\begin{equation}
{\dot M}_0\sim \pi r_{\rm acc}^2 v_{\rm rel}\rho(24R_c)\sim
5\times 10^{15}M_{p1}^2~{\rm g/s},
\label{dotM0}
\end{equation}
where we choose $v_{\rm rel}\sim \sqrt{(v_k-v_{\rm
peria})^2+v_s^2}\sim 50$ km/s for the following estimates.
The pulsar could then accrete $\Delta M\sim 9\times
10^{21}M_{p1}^2$ g for $\sim 20$ days near each
periastron passage. An effective accretion rate averaged over the
orbital period would then be ${\dot M}\simeq (20~{\rm days}/P_{\rm
orb}){\dot M}_0$.

The magnetic momentum can be estimated from $\Omega=2\pi/P$ and
$\dot \Omega$,
\begin{equation}
\mu\simeq ({4\pi c^3\over 5}{-{\dot\Omega}\over \Omega^3}{\bar
\rho} R_p^5)^{1/2}\sim 3\times 10^{29}M_{p1}^{5/6}~{\rm G~cm^3},
\label{mu}
\end{equation}
for a pulsar with radius $R_p$ and mass $M_p$ (averaged density
${\bar \rho}=3M_p/(4\pi R_p^3)\sim 4\times 10^{14}$ g/cm$^3$ for
quark stars with mass $<\sim M_\odot$).
The magnetospheric radius would then be
\begin{equation}
r_m\simeq (2GM_p)^{-1/7}\mu^{4/7}{\dot M}^{-2/7}\sim 1.2\times
10^{13}M_{p1}^{1/3}{\dot M}^{-2/7}~{\rm cm},
\label{rm}
\end{equation}
which is likely much larger than $r_{\rm lc}$, and is thus $\gg
r_{\rm co}$. A lower pulsar mass results in  a higher ratio of
$r_m/r_{\rm lc}$ (Fig. 1). This means that the accretion with a
rate of even ${\dot M}\simeq 5\times 10^{15}$ g/s can not drive
matter to reach the pulsar surface, but is in a propeller phase.
That $r_m\gg r_{\rm lc}$ could be a direct cause for PSR
B1259-63's being a {\em radio} pulsar.
\begin{figure}
  \centering
    \includegraphics[width=8cm]{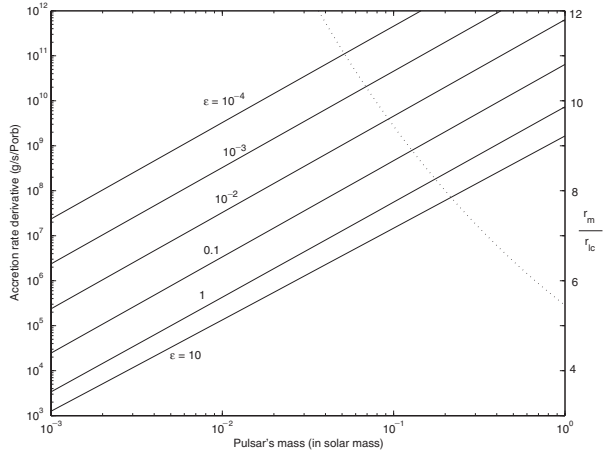}
    \caption{%
The accretion rate derivative (solid lines, in unit of ${\rm
g}\cdot {\rm s}^{-1}\cdot P_{\rm orb}^{-1}$), $\ddot M$, is
computed as a function of pulsar's mass in order for the observed
braking index to be -36.7 by Eq.(\ref{n}). Each curve is for a
different value of $\varepsilon$, the meaning of which can be
found in Eq.(\ref{dotO}). In the calculation, we choose an average
accretion rate of ${\dot M}=(20~{\rm days}/P_{\rm orb}){\dot
M}_0$, assuming the pulsar accretes for 20 days with a rate $\sim
{\dot M}_0$. The dotted line shows the magnetospheric radius $r_m$
in units of the light cylinder radius $r_{\rm lc}$. It is shown
that $r_m/r_{\rm lc}\simeq 5.5$ for $M_p=M_\odot$, but increases
as $M_p$ decreases.
\label{f1}}
\end{figure}

The magnetospheric interaction between a pulsar and its
surrounding disk is still not well understood. Menou et al. (2001)
discussed disk torque for the case of $r_{\rm m}\sim r_{\rm lc}$,
and recognized that the torque would likely be largely reduced or
entirely suppressed if $r_{\rm m}\gg r_{\rm lc}$.
Both the magnetodipole radiation and the propeller effect may
result in the pulsar's spindown,
\begin{equation}
{\dot \Omega}=-A\Omega^3-\varepsilon B {\dot M}^{3/7}\Omega,
\label{dotO}
\end{equation}
with
$$%
A=2\mu^2/(3c^3I), ~~B\simeq 2.88\times 10^{26}M_{p1}^{2/3}/I,
$$%
where $I=2M_pR^2/5$ is the moment of inertia, and the propeller
torque, ${\dot \Omega}=-2{\dot M}r_m^2\Omega/I$ of the second term
in Eq.(\ref{dotO}), proposed by Menou et al. (1999) is applied,
with a suppression factor introduced, $\varepsilon$. The
magnetodipole torque is generally smaller, but almost of the same
order, as the propeller torque for $\varepsilon=1$. In other
words, the magnetodipole torque is generally larger than the
propeller torque for $\varepsilon<0.1$. The calculation of $\mu$ in
Eq.(\ref{mu}) is therefore consistent.
One can also calculate the braking index from Eq.(\ref{dotO}),
\begin{equation}
n={\Omega{\ddot \Omega}\over {\dot \Omega}^2}=-{{\Omega}\over
{\dot \Omega}^2}[(3A\Omega^2+\varepsilon B{\dot M}^{3/7}){\dot
\Omega}+{3\over 7}\varepsilon B \Omega {\dot M}^{-4/7}{\ddot M}].
\label{n}
\end{equation}
Since $\dot \Omega<0$, it is evident that the braking index $n>0$
for an accretion with constant rate, $\ddot M=0$. However, the
observed index is negative, $n=-36.7<0$, which means the effective
accretion rate increases, $\ddot M>0$.

Since the accretion rate derivative, $\ddot M$, and the
suppression factor, $\varepsilon$, are two parameters hitherto
unknown, let's find the relationship between them in order to
explain the observed
 braking index $n=-36.7$.
The results, numerically calculated via Eq.(\ref{n}), are shown
in Fig. 1.
It is found that high $\varepsilon$ and/or low pulsar-mass ($M_p$)
would result in a small $\ddot M$. The averaged accretion rate of
${\dot M}=(20~{\rm days}/P_{\rm orb}){\dot M}_0\sim 8\times
10^{13}$ g/s. PSR B1259-63 could be a low-mass quark star (with
$M<0.1M_\odot$) if $\varepsilon \la 10^{-4}$ and $\ddot M<
10^{12}$ g/s/$P_{\rm orb}$.

The change of accretion rate on long timescales ($\ga P_{\rm orb}$)
was considered in the previous paragraph.
It is natural to suggest that the accretion rate also increases
before (but decays soon after) the periastron, e.g.,
exponentially, in a form of ${\dot M}\sim {\dot M}_0 e^{\pm
t/\tau}$ on a short timescale $\tau < P_{\rm orb}$.
This short-term variation of $\dot M$ could significantly affect
the timing behaviors close to the periastron, which could be the
reason that a few timing points have systematic discrepancies at
periastron passages.

\section{Conclusion and Discussion}

A few scenarios for understanding the $x$-jumps at the periastron
are presented, including planet-like objects around the Be star,
asymmetric mass-distribution by the Be star's oscillations, and a
density-wave pattern in the Be star's disk excited by objects
orbiting the Be star. The first one could be more likely than the
others.
We show that an accretion disk forms around the rapidly rotating,
strongly magnetized pulsar because of the circularization radius
being much larger than the radius of the pulsar's magnetospheric
radius. The inner region of the disk is also beyond the
light-cylinder, provided that the pulsar's accreted matter is
replenished in the Keplerian disk of companion SS 2883. Propeller
torque would therefore result in the pulsar's spindown, besides
that due to magnetodipole radiation, though it may be suppressed
by a factor $\varepsilon$.
The accretion rate should increase on a timescale of $\ga P_{\rm
orb}$ in order to explain the negative braking index $n=-36.7$,
but the reason for this increase is not certain (for example it may be due to a
secular changing of $a_p$ via tidal effects).
Timing models with the inclusion of disk propeller effects would
be necessary for fitting the timing data, and further observations
would  in turn check the model.

The PSR J0045-7319 binary system (Kaspi et al. 1996) is similar to
PSR B1259-63, in that tidal interactions may also be enhanced near
the periastron in that system (Witte \& Savonije 1999).
But why are their timing behaviors so different?
This could be due to
(1) the differences of pulsar masses (the mass of PSR J0045-7319
might be much lower than that of PSR B1259-63, $M_{1259}\ll
M_{0045}$), (2) the differences of companion masses
($M_{1259}^c>M_{0045}^c$), and (3) the different evolutionary
stages of the companions (a younger Be star may oscillate with
higher amplitude, or, planet-like objects could have been captured
or evaporated by an old B star).

A very negative braking index might be a feature of the disk with
increasing mass.
Long-term timing of AXP/SGRs and CCOs is proposed to see if there
are propeller disks around them (the index $n>0$ for a decaying
disk).
In addition, this work has implications for studying {\em
long}-term timing behavior of radio pulsars with propeller disks
formed from either supernova fall-back onto young neutron stars or
interstellar medium medium accretion onto old ones (probably near
the death line in the $P-{\dot P}$ diagram).


{\em Acknowledgments}:
The author thanks Dr. Na Wang, Dr. Yuqing Lou, and Prof. Yuefang
Wu for helpful discussions. I am in debt to Prof. Joel Weisberg
for his stimulating comments and suggestions, and improving the
language. I would like as well to thank Dr. Xiangdong Li for his
valuable comments which help me to improve the paper. This work is
supported by NSFC (10573002) and the Key Grant Project of Chinese
Ministry of Education (305001).



\begin{thebibliography}{}

\bibitem{} Boss, A. 2003, ApJ, 599, 577

\bibitem{} Bryden, G., Beichman, C. A., Rieke, G. H. 2006,
preprint (astro-ph/0604115)

\bibitem{} Chakraborty, A., Ge, J., Mahadevan, S. 2004, ApJ, 606, L69 (astro-ph/0403448)

\bibitem{} Chatterjee, P., Hernquist, L., Narayan, R. 2000, ApJ, 534, 373

\bibitem{} Frank, J., King, A., Raine, D. 2002, {\em Accretion power in
astrophysics} (3rd Edition), Cambridge Univ. Press, \S4.9

\bibitem{} Fuente A., et al. 2003, ApJ, 598, L39 (astro-ph/0310062)

\bibitem{} Johnston, S., Ball, L., Wang, N., Manchester, R. N.
2005, MNRAS, in press (astro-ph/0501660)

\bibitem{} Johnston S., et al. 1992, MNRAS, 255, 401

\bibitem{} Kaspi V. M., et al. 1996, Nature, 381, 584

\bibitem{} Kochanek, C. S. 1993, ApJ, 406, 638

\bibitem[Lipunov 1992]{lip92}
Lipunov V. M. 1992, Astrophyics of Neutron Stars, Springer-Verlag:
Berlin

\bibitem{} Manchester R. N., et al. 1995, ApJ, 445, L137

\bibitem{} Mardling, R. A. 1995, ApJ, 450, 722

\bibitem{} Livingstone, M. A., Kaspi, V. M., Gotthelf, E. V. 2006,
preprint (astro-ph/0601530)

\bibitem{} Menou, K., Esin, A. A., Narayan, R., Garcia, M. R., Lasota, J.-P.,
Mc-Clintock, J. E. 1999, ApJ, 520, 276

\bibitem{} Menou, K., Perna, R., Hernquist, L. 2001, ApJ, 554, L63

\bibitem{} Tremaine, S., Zakamska, N. L. 2004, in: The search for other worlds:
Fourteenth Astrophysics Conference, AIP Conf. Proc. v.713, eds. S.
S. Holt \& D. Deming, p.243 (astro-ph/0312045)

\bibitem{} Wang, N., Johnston, S., Manchester, R. N. 2004, MNRAS,
351, 599

\bibitem{} Wang, Z., Chakrabarty, D., Kaplan, D. L. 2006, Nature,
440, 775 (astro-ph/0604076)

\bibitem{}  Wex N., et al. 1998, MNRAS, 298, 997

\bibitem{} Witte M. G., Savonije G. J. 1999, A\&A, 350, 129

\bibitem{} Xu, R. X. 2005, MNRAS, 356, 359

\bibitem{} Xu, R. X., Qiao, G. J. 2001, ApJ, 561, L85

\bibitem{} Xu, R. X, Wang, H. G., Qiao, G. J., 2003, Chin. Phys. Lett., 20, 314


\end{thebibliography}
\end{document}